\documentclass[aps,prb,amsmath,amssymb,twocolumn,a4paper,floatfix]{revtex4-1}
\usepackage{graphicx}
\usepackage{float}
\usepackage{hyperref}
\usepackage{color}

\begin{document}
\title{Finite-temperature phase transitions in the ionic Hubbard model }
\author{Aaram J. Kim}
\affiliation{Department of Physics and Astronomy and Center for Theoretical Physics, Seoul National University, Seoul 151-747, Korea}
\author{M. Y. Choi}
\affiliation{Department of Physics and Astronomy and Center for Theoretical Physics, Seoul National University, Seoul 151-747, Korea}
\author{Gun Sang Jeon}
\email{gsjeon@ewha.ac.kr}
\affiliation{Department of Physics, Ewha Womans University, Seoul 120-750, Korea}

\begin{abstract}
We investigate paramagnetic metal-insulator transitions in the infinite-dimensional ionic Hubbard model at finite temperatures.
By means of the dynamical mean-field theory with an impurity solver of the
continuous-time quantum Monte Carlo method, we show that an increase in the
interaction strength brings about a crossover from a band insulating phase to a metallic one,
followed by a first-order transition to a Mott insulating phase.
The first-order transition turns into a crossover above a certain critical
temperature, which becomes higher as the staggered lattice potential is increased.
Further, analysis of the temperature dependence of the energy density discloses that the
intermediate metallic phase is a Fermi liquid.
It is also found that the metallic phase is stable against
strong staggered potentials even at very low temperatures.
\end{abstract}


\maketitle

\section{Introduction}
\label{sec:intro}
The effects of correlations between electrons have been one of the most fascinating
topics in modern condensed matter physics.
A variety of remarkable phenomena such as superconductors with high critical
temperatures~\cite{Dagotto1994} and interaction-driven metal-insulator
transitions~\cite{Imada1998} is well known to arise from electron correlations.
In describing such electron correlations, the Hubbard model (HM) opened a new paradigm.
It has proved to be successful in capturing the essential physics of 
correlation-induced phenomena by incorporating just a few simple ingredients:
tight-binding electrons with the local Coulomb interaction.
Interesting variants of the HM have been proposed
to investigate correlation effects in the band insulator (BI).
One of the popular examples is the ionic Hubbard model (IHM), where tight-binding
electrons interact via the local Coulomb interaction under a staggered lattice
potential.~\cite{Hubbard1981,Torrance1981b,Torrance1981a}
It was first applied to the study of the neutral-ionic transition in a
charge-transfer organic chain~\cite{Nagaosa1986a,Avignon1986,Luty1987,Yonemitsu2002,Wilkens2001,Lemee-Cailleau1997,Caprara2000a,Caprara2000,Horiuchi2000,Anusooya-Pati2001} and also suggested as a model
for the polarization phenomena of ferroelectric perovskite
materials~\cite{Egami1993,Ishihara1994,Resta1995,Ortiz1996,Gidopoulos2000,Torio2001} and Kondo insulators such as
FeSi and FeSb$_2$.~\cite{Kunes2008}

On a bipartite lattice, the staggered lattice potential of the IHM
doubles the periodicity of the system, giving rise to a gap at the zone boundary.
Accordingly, in the noninteracting limit the system prefers a band insulating
phase where most electrons stay on a sublattice with lower potential.
The resulting BI competes with a Mott insulator (MI) with one electron per lattice site,
which is driven by local interactions.
This competition is expected to enrich the physics in the transition between the two phases,
which has been studied theoretically for decades.

The emergence of an intermediate phase has been studied in
one dimension~\cite{Luty1987,Fabrizio1999,Wilkens2001,Zhang2003,Manmana2004,Kampf2003,Go2008,Go2011,Yonemitsu2002,Pozgajcic2003,Aligia2005,Aligia2004a,Fabrizio1999,Fabrizio2000,Japaridze2007,Batista2004,Torio2006,Manmana2004,Kampf2003,Lou2003,Tincani2009,Otsuka2005,Zhang2003} and in two dimensions.~\cite{Paris2007, Bouadim2007, Kancharla2007a, Chen2010}
In one dimension, it was revealed by the bosonization method that a spontaneously dimerized insulating phase shows up between the BI and the MI,~\cite{Fabrizio1999,Fabrizio2000,Japaridze2007,Batista2004,Torio2006} which was confirmed subsequently in numerical studies.~\cite{Manmana2004,Kampf2003,Lou2003,Tincani2009,Otsuka2005,Zhang2003}
Some peculiar spectral properties such as spin-charge separation were als
studied by the cellular dynamical mean-field theory (DMFT).~\cite{Go2008,Go2011}
Extensive investigations have also been made into the effects of additional degrees of freedom on the one-dimensional IHM, including electron-lattice coupling,~\cite{Avignon1986,Luty1987,Caprara2000a} spin-density wave,~\cite{Tugushev1996,Caprara2000,Pozgajcic2003,Caprara2000a} next-nearest-neighbor interaction,~\cite{Avignon1986,Luty1987,Caprara2000,Yonemitsu2002,Aligia2005,Torio2001} asymmetry in electron hopping,~\cite{Yonemitsu2002,Wilkens2001} alternating Hubbard interaction,~\cite{Pozgajcic2003} periodicity of the lattice,~\cite{Torio2006} coupling with conducting leads,~\cite{Aligia2004} and next-nearest neighbor hopping.~\cite{Japaridze2007}
As to the nature of the intermediate phase in two dimensions, there is some controversy:
The determinant quantum Monte Carlo study~\cite{Paris2007, Bouadim2007}
predicted a metallic phase, while an insulating
phase was observed via the cellular DMFT or the variational cluster approach.~\cite{Kancharla2007a, Chen2010}

In infinite dimensions, on the other hand, the single-site DMFT has revealed two successive
metal-insulator transitions at zero temperature.~\cite{Garg2006,Craco2008,Byczuk2009a,Lombardo2006}
Weak interactions tend to reduce the single-particle gap, driving the system
into a metallic phase.
The system eventually becomes an MI, caused by the further increase in the
interaction strength.
Here it is remarkable that a metallic phase emerges due to correlation effects
of Coulomb interactions;
this is in sharp contrast with the intermediate insulating phase,
which is confirmed in the one-dimensional IHM.
The effects of antiferromagnetic ordering induced by local
interactions have also been studied in the IHM.~\cite{Byczuk2009a,Wang2014}

In this paper, we focus on the finite-temperature properties of the transitions
between paramagnetic phases in the infinite-dimensional IHM at half-filling.
We adopt the DMFT combined with the continuous-time quantum Monte
Carlo (CTQMC) method.~\cite{Rubtsov2005,Werner2006,Gull2008,Gull2011}
First, the spectral properties of the IHM are examined at finite temperatures.
The Fermi-level spectral weight, which can be estimated
from the imaginary-time Green function, demonstrates that with an increase
in the local interaction the system exhibits a crossover from BI to metal,
which is followed by a discontinuous transition to an MI.
The spectral function as well as local quantities such as double occupancy and
staggered charge also supports the above description of the transition behaviors.
The energy density, which can be measured directly from the CTQMC method, shows that
the metallic phase always has a lower energy than the Mott insulating phase within the
coexistence region as in the standard HM.
The resulting finite-temperature phase diagram illustrates that the
crossover interaction strength between metal and MI decreases with the temperature.
It is also found that the metal-MI transition is similar to that in the HM while
the critical temperature tends to increase as the staggered lattice potential
becomes stronger.
The dependence of the total energy density on the temperature indicates that the
correlation-driven metallic phase is a Fermi liquid.
The phase diagram at very low temperatures shows that the metallic phase persists
for very strong staggered lattice potentials.

This paper is organized as follows:
In Sec.~\ref{sec:modelmethod} we introduce the IHM and describe how to deal with the model
by the single-site DMFT with the CTQMC as an impurity solver.
Section~\ref{sec:results} presents the results of our numerical calculations.
We examine spectral properties, local quantities, and several components of energy densities,
based on which the phase diagram is constructed.
We also investigate the nature of the intermediate metallic phase and the
dependence of the transition on the strength of the staggered lattice potential.
Finally, we conclude the paper by summarizing the results in
Sec.~\ref{sec:summary}.

\section{MODEL AND METHODS}
\label{sec:modelmethod}
We consider the IHM on a bipartite lattice, the Hamiltonian of which is given by
\begin{eqnarray}
	\mathcal{H} &=& -t \sum_{\langle ij\rangle\sigma}
	(	\hat{c}_{j\sigma}^{\dagger}\hat{c}_{i\sigma}
	+ \hat{c}_{i\sigma}^{\dagger}\hat{c}_{j\sigma} )
	+ U\sum_{i}\hat{n}_{i\uparrow}\hat{n}_{i\downarrow}~\nonumber\\
	& & + \sum_{i\sigma}\epsilon_i \hat{n}_{i\sigma} - \mu\sum_{i\sigma} \hat{n}_{i\sigma},
\end{eqnarray}
where $\hat{c}_{i\sigma}$/$\hat{c}^{\dagger}_{i\sigma}$ is the
annihilation/creation operator of an electron with spin
$\sigma$ at the $i$th lattice site. The corresponding number operator is defined to be
$\hat{n}_{i\sigma} \equiv \hat{c}^{\dagger}_{i\sigma}\hat{c}_{i\sigma}$.
The parameters $t$ and $U$ represent the nearest-neighbor hopping amplitude and the
Hubbard interaction, respectively.
The lattice is a bipartite one composed of two sublattices, $A$ and $B$,
and the local lattice potential energy $\epsilon_i$ is given by
\begin{equation}
	\epsilon_i = \left\{
		\begin{array}{ll}
			\Delta & \hbox{for } i \in \textrm{A},
			\\
			- \Delta & \hbox{for } i \in \textrm{B}.
		\end{array}
		\right.
\end{equation}

In this work we adopt the single-site DMFT, which is exact in infinite
dimensions.~\cite{Georges1996}
Within the DMFT, the original lattice model is mapped onto a single-impurity
Anderson model, which is described by the Hamiltonian
\begin{eqnarray}
	\mathcal{H}^{\alpha}_\textrm{SIAM} &=& (\varepsilon_\alpha - \mu )\hat{n}_{\alpha\sigma} + \sum_k ( V_{k\alpha\sigma}\hat{c}^{\dagger}_{\alpha\sigma}\hat{a}_{k\sigma} + h.c.)~\nonumber\\
	& &+ U\hat{n}_{\alpha\uparrow}\hat{n}_{\alpha\downarrow} + \sum_k \varepsilon_{k\sigma}\hat{a}^{\dagger}_{k\sigma}\hat{a}_{k\sigma}.
\end{eqnarray}
Here $\hat{c}_{\alpha\sigma}$/$\hat{c}^{\dagger}_{\alpha\sigma}$ is the
annihilation/creation operator of an electron at the impurity corresponding to
sublattice $\alpha$, and
$\hat{a}_{k\sigma}$/$\hat{a}^{\dagger}_{k\sigma}$ is the annihilation/creation
operator of an electron at the $k$th bath site which has on-site energy
$\varepsilon_{k\sigma}$ and is coupled with the impurity via the hybridization
matrix element $V_{k\alpha\sigma}$.

The structure of a bipartite lattice leads to an impurity Green function of the form
\begin{equation}\label{eq:hilbert}
	G_{\alpha}(i\omega_n) = \zeta_{\bar{\alpha}}\int_{-\infty}^{\infty} d\varepsilon \hspace{5pt} \frac{\rho_0({\varepsilon})}{\zeta_{\alpha}\zeta_{\bar{\alpha}} - \varepsilon^2}
\end{equation}
for $(\alpha, \bar{\alpha}) = (A, B)$ and $(B, A)$,
where $\rho_0(\varepsilon)$ is the bare density of states (DOS) of the lattice and
$\zeta_{\alpha} \equiv i\omega_n -\varepsilon_\alpha + \mu - \Sigma_{\alpha}(i\omega_n)$
with the self-energy $\Sigma_{\alpha}$ and Matsubara frequency $\omega_n$.
The calculation is performed on the Bethe lattice, where the DOS is given in the semicircular form:
$\rho_{0}(\varepsilon) = (2/\pi D)\sqrt{1 - (\varepsilon/D)^2}$.
Through this paper we use the half-band width $D=2t$ as the unit of energy.

The DOS of a semicircular form allows analytic integration of Eq.~\eqref{eq:hilbert},
which yields
\begin{equation}
G^{-1}_{\alpha}(i\omega_n) = \zeta_{\alpha} -
\frac{D^2}{4}G_{\bar{\alpha}}(i\omega_n).
\end{equation}
With the help of the particle-hole symmetry, we have the following relations:
\begin{eqnarray}
\Sigma_{\alpha}(i\omega_{n}) &=& U - \Sigma_{\bar{\alpha}}(-i\omega_n), \nonumber \\
G_{\alpha}(i\omega_n) &=& -G_{\bar{\alpha}}(-i\omega_n).
\end{eqnarray}
Then the Dyson's equation, $\mathcal{G}_{0\alpha}^{-1} =
\Sigma_\alpha(i\omega_n) + G^{-1}_{\alpha}(i\omega_n)$, reduces to
\begin{equation}
	\mathcal{G}_{0\alpha}^{-1}(i\omega_n) = i\omega_n - \varepsilon_\alpha +
	\mu + \frac{D^2}{4}G_{\alpha}(-i\omega_n),
	\label{eq:integration}	
\end{equation}
which imposes the self-consistency relation on the impurity problem.

We solve the impurity problem only in sublattice $A$ to obtain
$G_A(i\omega_n)$ from $\mathcal{G}_{0A}(i \omega_n)$ by means of the CTQMC method based
on the hybridization expansion, which has proven to be efficient particularly in
the strong-interaction regime.
We typically use $10^8$ Monte Carlo steps for each DMFT iteration, which turns
out to be sufficient to achieve the required accuracy of the Green function at the
lowest temperature, $T=1/128$.
The self-consistency loop is iterated 50 times for the convergence of the
solution within the DMFT.

\section{Results}
\label{sec:results}
\subsection{Spectral properties}

\begin{figure}
	\includegraphics[width=0.4\textwidth]{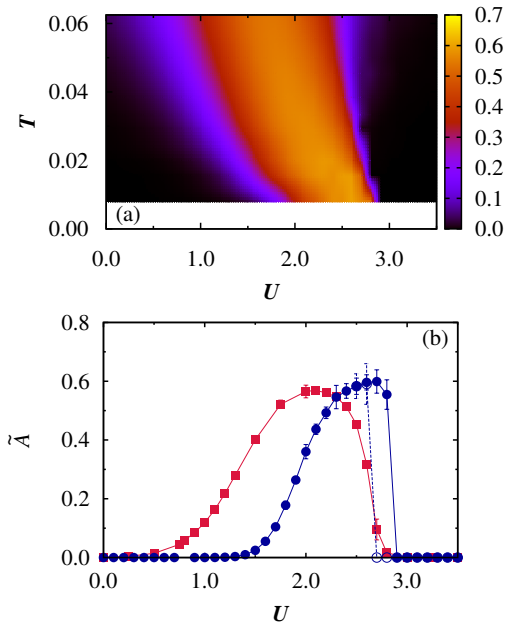}
	\caption{(Color online) Fermi-level spectral weight
		$\tilde{A}_{\alpha}$ for $\Delta = 0.5$.
	  (a) The colored plot displays $\tilde{A}_\alpha$
      on the plane of the interaction strength $U$ and temperature $T$,
      obtained via increasing $U$.
	  (b) $\tilde{A}_\alpha$ at temperatures $T=1/32$ [(red) squares] and
	  $1/128$ [(blue) circles]. Filled and open symbols for $T=1/128$ represent
      data obtained via increasing and decreasing $U$, respectively.
}
\label{fig:A0}
\end{figure}

To probe the metal-insulator transitions, we consider the Fermi-level spectral weight:
\begin{eqnarray}
	\tilde{A}_\alpha &\equiv& -\frac{1}{\pi T}
	G_\alpha (\tau{=}1/2T) \nonumber\\
		   &=& \frac{1}{2\pi T}\int_{-\infty}^{\infty}~d\omega
	\frac{1}{\cosh(\omega/2T)}A_\alpha(\omega),
\end{eqnarray}
where 
$
A_{\alpha}(\omega) \equiv - (1/\pi) \textrm{Im} G_\alpha (\omega {+} i0^+)
$ is the spectral function of sublattice $\alpha$.
At very low temperatures $\tilde{A}_\alpha$ is approximately the same as the Fermi-level spectral function
$A_\alpha (\omega{=}0)$.
Since the imaginary-time Green function can be measured directly from Monte Carlo
sampling, it is frequently used to examine the metal-insulator
transition.~\cite{Trivedi1995,Gull2008,Gull2010}

In Fig.~\ref{fig:A0}, we present the Fermi-level spectral weight as
a function of the temperature $T$ and the interaction strength $U$.
The colored plot on the plane of $U$ and $T$ clearly demonstrates that
two insulating phases (dark regions) are separated by an intermediate
metallic phase (bright region).
As clarified in the existing zero-temperature studies,~\cite{Garg2006,Craco2008,Byczuk2009a}
the insulating phase for weak interactions correspond to a BI, while that at strong
interactions represents an MI.

The BI connects smoothly with the metallic phase via a finite-width crossover
region at finite temperatures.
As the temperature is lowered, the onset value of $\tilde{A}$ becomes steeper
and the size of the crossover region decreases appreciably;
this is consistent with the continuous transition observed at zero temperature.

For strong interactions,
on the other hand, we observe a rather steeper transition between the metal and
the MI at finite temperatures.
Below a certain critical temperature, the Mott transition turns out to be of 
first order, which is evidenced by
the presence of the hysteretic behavior displayed at $T=1/128$ in
Fig.~\ref{fig:A0}(b).
Accordingly, we have
lower and upper transition interaction strengths, $U_{c1}$ and $U_{c2}$, at
which MI and metallic phases become unstable, respectively.
Thermodynamic phase transitions occur between $U_{c1}$ and $U_{c2}$ at finite
temperatures; the determination of the phase transition line is discussed
later.
Above the critical temperature, the boundary between the metal and the MI also
appears as a crossover, and the crossover region expands as the temperature is increased.

\begin{figure}
		\includegraphics[width=0.4\textwidth]{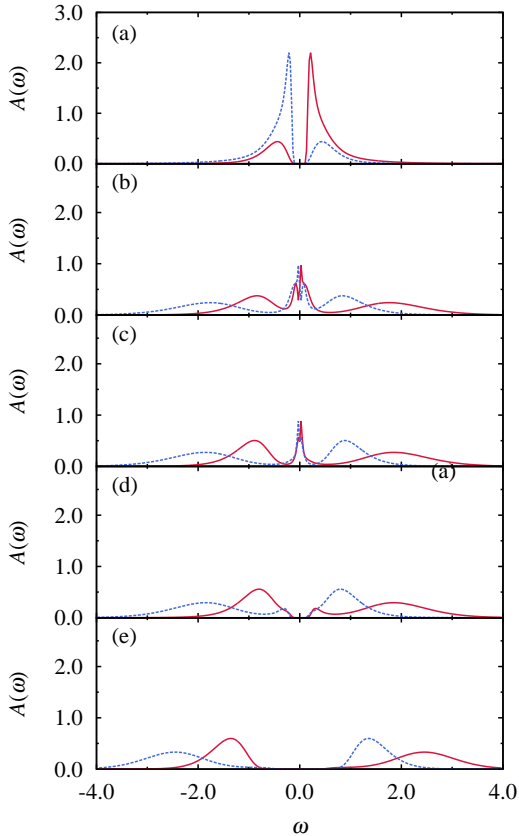}
	\caption{(Color online)
	Spectral function $A(\omega)$ for $\Delta=0.5$ and $T=1/128$. Corresponding interaction strengths are (a) $1.0$, (b) $2.5$, (c) $2.8$ (increasing $U$), (d) $2.8$ (decreasing $U$), and (e) $4.0$. Solid (red) and dotted (blue) lines represent the spectral function at sublattices $A$ and $B$, respectively.}	
\label{fig:MEM}
\end{figure}

We use the maximum entropy method (MEM) for analytic continuation of the
Matsubara Green function to the real frequency domain and obtain the spectral
function $A(\omega)$.
The resulting spectral function is presented in Fig.~\ref{fig:MEM}.
In the region of weak interactions, the single-particle gap is formed
around the Fermi level with singular behavior at the band edge,
which is reminiscent of the noninteracting DOS with a van Hove singularity.
We also observe that in the occupation of each sublattice there is a significant imbalance between $A$ and $B$ sublattices,
which is a characteristic feature of the BI.

On the other hand, the Mott gap emerges with a prominent four-peak
structure for strong interactions.
For a given sublattice, two peaks correspond to the upper and the lower Hubbard
bands, respectively.
The upper or lower Hubbard bands on different sublattices are separated by the
staggered lattice potential $\Delta$.
Both Hubbard bands on sublattice $B$, having the lower lattice potential,
are located at a lower energy compared with those on sublattice $A$.

In the intermediate-interaction region,
we observe a metallic phase with a finite spectral weight at the Fermi level.
In this phase a quasiparticle peak near the Fermi level is surrounded by
four Hubbard bands, and the disappearance of the quasiparticle peak signifies
the onset of a Mott phase.
The quasiparticle peak also shows pseudogap-like behavior around the
Fermi level, which is discussed in the zero-temperature study.~\cite{Craco2008}
At temperature $T=1/128$,
there exists a coexistence region where both metal and MI are locally stable.
Figures~\ref{fig:MEM}(c) and ~\ref{fig:MEM}(d) correspond to metallic and
insulating solutions, respectively.
The overall features of the spectral functions are in good agreement with the
previous zero-temperature results obtained via NRG.~\cite{Byczuk2009a}

Before going on to the next section, we make some comments on the stability of our MEM procedure.
The stability investigation shows that our MEM procedure is reliable enough
to characterize the fine structures of spectral function.
For example, the pseudogap-like behavior around the Fermi level
in Fig.~\ref{fig:MEM}(b) 
is robust against the statistical
fluctuations of the imaginary-time Green function. 
In our calculations
the statistical error of the imaginary-time Green function is around
order $10^{-4}$.
We have also checked the stability of the MEM procedure by examining the dependence
on the model function and the scaling parameter selection, 
which turns out to have negligible effects on the resulting spectral function.

\begin{figure}
	\includegraphics[width=0.4\textwidth]{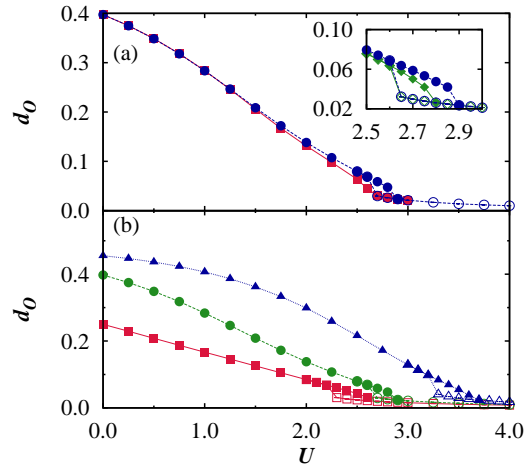}
	\caption{(Color online) Double-occupancy $d_O$ as a function of $U$:
	(a)~for staggered lattice potential $\Delta=0.5$ at temperatures $T=1/32$ [(red) squares] and $T=1/128$ [(blue) circles] and
	(b)~at temperature $T=1/128$ for various values of the staggered lattice potential-from top to bottom, $\Delta =1.0, 0.5$, and $0.0$.
	The inset in (a): Detailed behavior in the coexistence region. 
	Data for $T=1/64$, (green) diamonds; $T=1/128$, (blue) circles. 
	Data obtained via increasing $U$, filled symbols; via decreasing $U$, open symbols.
}
	\label{fig:do}
\end{figure}

\begin{figure}
	\includegraphics[width=0.4\textwidth]{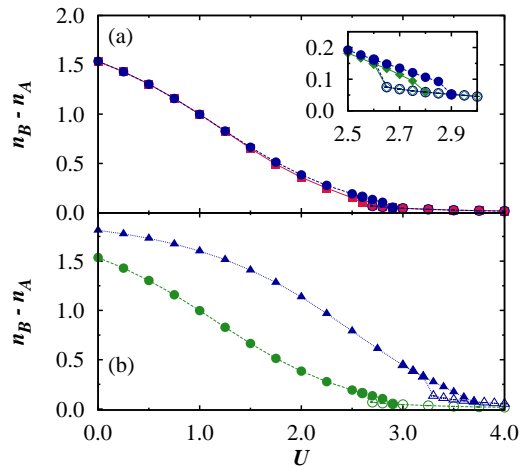}
	\caption{(Color online) Staggered charge density, $n_B-n_A$ as a function of $U$: 
		(a)~for staggered lattice potential $\Delta=0.5$ at temperatures $T=1/32$ [(red) squares] 
		and $T=1/128$ [(blue) circles] and
		(b)~at $T=1/128$ for $\Delta =0.5$ [(green) circles] and $1.0$ [(blue) triangles]. 
		Inset in (a): Details in the coexistence region.  Data for $T=1/64$, (green) diamonds; $T=1/128$, (blue) circles. 
		Data obtained via increasing $U$, filled symbols; decreasing $U$, open symbols.
	}
	\label{fig:scd}
\end{figure}

\subsection{Local quantities}
The staggered charge density is given by the difference between the number
densities at two sublattices, $n_A{-}n_B$, with the sublattice number density
defined to be
$ n_\alpha \equiv \sum_\sigma \langle \hat{n}_{\alpha \sigma} \rangle $
for $\alpha = A$ and $B$.
We also compute the double occupancy $d_O$ given by
\begin{equation}
d_O \equiv \frac12\sum_{\alpha} \langle \hat{n}_{\alpha\uparrow} \hat{n}_{\alpha\downarrow} \rangle .
\end{equation}

The results for the double occupancy and the staggered charge density are
shown in Figs.~\ref{fig:do} and \ref{fig:scd}.
In the IHM, the interaction strength $U$ competes with the staggered lattice
potential $\mathit{\Delta}$ due to different favorable electron configurations.
While the staggered lattice potential forces electrons to stay at the lower
potential sites on sublattice $B$, the interaction, giving rise to energy cost,
tends to prevent two electrons from occupying the same site.
In the weak-interaction region, electrons prefer to gather on sublattice $B$ and
the system experiences an imbalance between the two sublattices, resulting in a
higher double occupancy, compared with the HM, corresponding to
$\Delta =0$, and a nonzero staggered charge density.
Such tendencies become stronger as $\Delta$ grows.

As the interaction strength is increased, both the double-occupancy and the
staggered charge density decrease monotonically with the imbalance between the two
sublattices becoming weaker.
In the MI phase, the staggered charge density is close to $0$.
However, the sublattice symmetry is broken in the Hamiltonian of the IHM
and the staggered charge density does not exactly vanish for any finite $U$.

In the coexistence region, the metallic phase always exhibits higher values of
the staggered charge density and double-occupancy than those in the MI phase.
The data at two temperatures, $T = 1/64$ and $1/128$, are compared
in the insets in Figs.~\ref{fig:do} and \ref{fig:scd}.
It is observed that the coexisting region widens as the temperature is lowered.
Further, the critical interaction strength is shown to increase with the staggered lattice potential.

\subsection{Energy density}

\begin{figure}
	\includegraphics[width=0.4\textwidth]{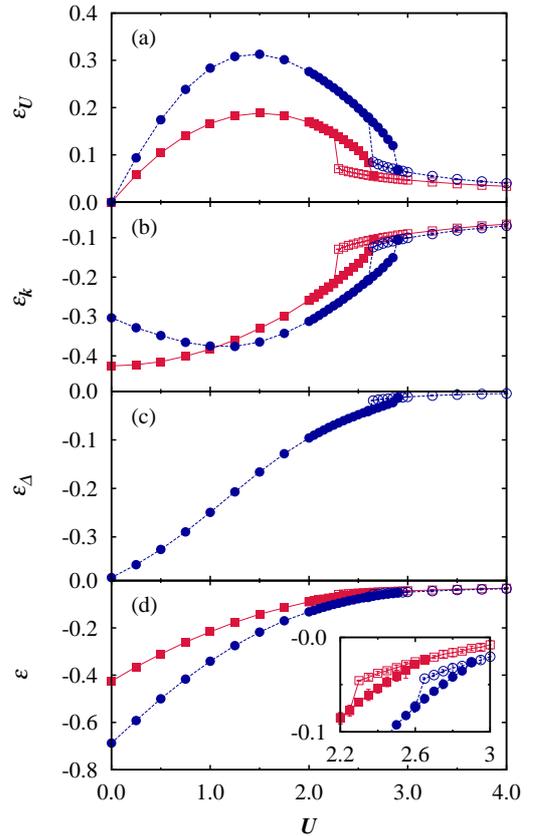}
	\caption{(Color online) Energy densities as functions of $U$ at temperature $T=1/128$ and $\Delta=0.5$. 
		(a) Interaction energy $\varepsilon_U$, (b) kinetic energy $\varepsilon_k$, 
		(c) staggered lattice potential energy $\varepsilon_{\Delta}$, and (d) total energy $\varepsilon$ (see text for definitions). 
		The (blue) circles and (red) squares represent data for the IHM and the HM, respectively.
		Inset in (d): Total energy density in the coexisting region; 
		DMFT solutions obtained via increasing $U$ (filled symbols) and decreasing $U$ (open symbols).
	}
	\label{fig:energy}
\end{figure}

Here we attempt to analyze the competition of the phases in terms of energy densities.
At finite temperatures the free energy will also have the contribution of the
entropy. 
We expect that the energy analysis given below is still valid for explaining 
the qualitative behaviors at low temperatures considered.
The DMFT solution gives the total, kinetic, lattice potential, and interaction energies per site
in the forms:
\begin{eqnarray} \label{eq:energy}
	\varepsilon &=& \varepsilon_k + \varepsilon_\Delta + \varepsilon_U, \nonumber \\
	\varepsilon_k &=& \frac{T}{2} \sum_{\alpha\sigma} \langle k_{\alpha\sigma}\rangle,~\nonumber\\
	\varepsilon_\Delta &=& \frac{\Delta}{2} ( n_A - n_B ),~\nonumber\\
	\varepsilon_U &=& U d_O, 
\end{eqnarray}
where $\langle k_{\alpha\sigma}\rangle$ is the average perturbation order of the
spin $\sigma$ electron at the impurity of sublattice $\alpha$. This can
be directly measured from CTQMC simulations.~\cite{Werner2006,Haule2007,Kim2014}

In Fig.~\ref{fig:energy} we plot all four energy densities for $\Delta=0.5$,
together with those in the HM.
Comparison between HM and IHM results indicates that for weak interactions,
the gain in the staggered lattice potential energy
exceeds the sum of the loss of both interaction and kinetic energies.
In consequence, the total energy of the IHM is lower than that of the HM,
which agrees with the characteristic behavior of the BI.

In the metallic region,
the kinetic and interaction energies of the IHM behave qualitatively the same as
those of the HM.
Quantitatively, the kinetic energy of the IHM is, in general, lower than that of
the HM with the same interaction strength.
We also observe that the kinetic energy increases with the interaction strength $U$,
which is in sharp contrast to the generally decreasing behavior in the BI.
Such different behaviors of the metal and the BI give rise to
a minimum of the kinetic energy at the interaction strength which generally coincides with
the boundary between the BI and the metal.

In the MI phase, the staggered lattice potential energy becomes negligible;
as a result, the total energy difference between the HM and the IHM decreases significantly
and monotonically as the interaction strength $U$ is increased.
At the boundary between the MI and the metal, a first-order Mott transition is also present in the IHM.
The critical interaction strength increases when the staggered lattice potential is introduced.
In the coexistence region, the total energy density in the metallic phase is always lower than
that in the MI phase, which also holds in the case of the HM.~\cite{Georges1996}
We expect that at zero temperature the IHM also undergoes a continuous phase transition
between the MI and the metal at the critical strength $U_{c2}$ .

\subsection{Finite-temperature phase transition}

\begin{figure}
	\includegraphics[width=0.4\textwidth]{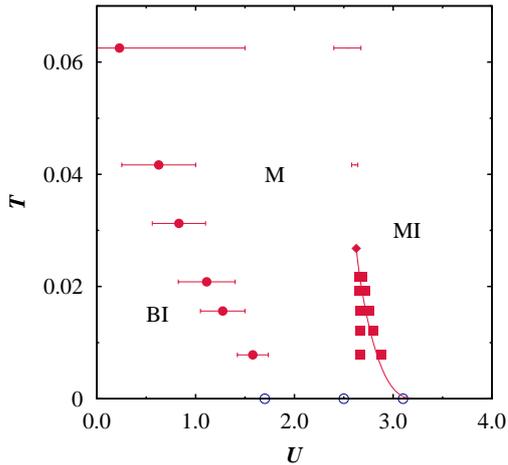}
	\caption{(Color online) Phase diagram for $\Delta=0.5$ on the plane of
		$T$ and $U$.
		Filled (red) circles and surrounding horizontal bars
		indicate the crossover strength $U_{co}$ and estimated crossover regions, respectively.
		Two transition points for the Mott transition, $U_{c1}$ and
		$U_{c2}$, are plotted by squares for various temperatures.
		The critical point of the Mott transition is represented by
		diamonds, along with the first-order transition line.
		Regions of the band insulator (BI), metal (M), and Mott insulator (MI) phases.
		The three open (blue) circles on the horizontal axis correspond to $U_{co}$, $U_{c1}$, and
		$U_{c2}$, respectively, obtained from NRG-DMFT at zero temperature
		(Ref.~\onlinecite{Byczuk2009a}).
	}
	\label{fig:TUphase}
\end{figure}

Based on the spectral properties as well as the local quantities, we may now
construct the phase diagram of the IHM.
Figure~\ref{fig:TUphase} exhibits the phase diagram for $\Delta=0.5$
on the plane of the temperature $T$ and the interaction strength $U$.
There exist three phases: metal, BI, and MI.
The BI and metal are connected through a crossover region while a
first-order Mott transition separates the metal from the MI.

As shown in Fig.~\ref{fig:A0}(b), the onset of $\tilde{A}$ becomes steeper as the
temperature is lowered.
Accordingly, at zero temperature, the transition between the BI and the metal is expected to
be continuous with a kink in $\tilde{A}$.
In order to estimate the crossover interaction strength $U_{co}$ at low
temperatures, we obtain a best linear fit of the area in which $\tilde{A}$
grows rather linearly in the metallic region.
We then estimate $U_{co}$ by the intersection point of the fitting line and
$\tilde{A}{=}0$.
The half-width of the crossover region is also identified as the distance from
$U_{co}$ to the linear region.
As the temperature is raised, the resulting $U_{co}$ tends to decrease and the width of the crossover-region increases.
It is also notable that $U_{co}$ estimated via CTQMC-DMFT in this work
gradually approaches the zero-temperature value obtained via NRG-DMFT.~\cite{Byczuk2009a}

At low temperatures we observe the coexistence region of the BI and metal
between $U_{c1}$ and $U_{c2}$, which can be identified by spectral functions
and local quantities such as double-occupancy and staggered charge densities.
With an increase in the temperature, $U_{c1}$ and $U_{c2}$ become closer, and the
coexistence region ceases to exist at a certain critical temperature, above
which the transition between the MI and the metal also appears as a crossover.
These general features are rather similar to those of the Mott transition in the HM.
Further, the extrapolation of $U_{c1}$ and $U_{c2}$ to zero temperature is
quite consistent with that of NRG-DMFT results.~\cite{Byczuk2009a}

By solving the differential equations constructed from the free-energy analysis,
we can obtain the first-order phase transition line, which is denoted by
the solid line in Fig.~\ref{fig:TUphase}.  
Using the thermodynamic relation
\begin{equation}
	\frac{\partial(\beta f)}{\partial\beta}\Bigg\vert_{U} 
	= \varepsilon~,
	\label{eq:diff_bf}
\end{equation}
we construct the differential equation of the interaction strength $U_c$ of the
first-order transition as a function of $T$, 
\begin{equation}
	\frac{dU_c(T)}{dT}= \frac{\delta \varepsilon(T,U)}{T\delta d_O(T,U)}~,
	\label{eq:diffeq}
\end{equation}
with $\Delta$ being fixed.
Here $f$ is the free-energy density and $\delta \varepsilon$ and $\delta d_O$
are the differences in the energy and the double-occupancy between metal and MI
in the coexistence region, respectively.
The numerical integration of Eq.~\eqref{eq:diffeq} gives the first-order
transition line.
The CTQMC procedure has the advantage that one can obtain the quantities necessary for the
differential equations directly from Monte Carlo sampling without any further approximation.
The details of the method can be found in Ref.~\onlinecite{Kim2014}, where the
HM is investigated by the same method.
The resulting transition line is plotted by the solid line in Fig.~\ref{fig:TUphase}.
The phase transition point at zero temperature is very close to $U_{c2}$ obtained from NRG-DMFT,
implying that the transition is continuous at zero temperature;
this is also the case in the HM without a staggered lattice potential.

\subsection{Nature of the intermediate metallic phase}

\begin{figure}
	\includegraphics[width=0.4\textwidth]{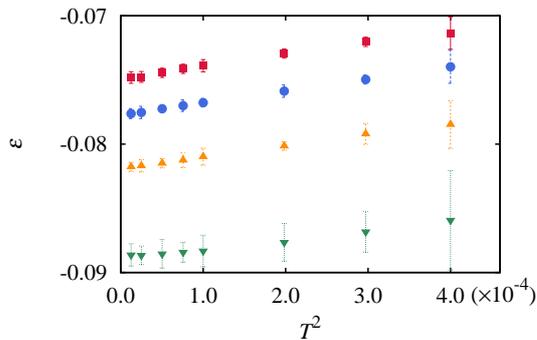}
	\caption{(Color online) Total energy density $\varepsilon$ as a
		function of temperature $T$. From top to bottom, the staggered lattice potential
		and interaction strength are given by
		$(\Delta, U) = (0, 2.2)$ (squares), $(0.5, 2.5)$ (circles), $(1, 3.2)$ (triangles), and $(3, 6.9)$ (inverted triangles).
		The horizontal axis is drawn on the scale of $T^2$.
	}
	\label{fig:FL}
\end{figure}

One interesting issue is the nature of the metallic phase present in the region of
intermediate interaction strengths.
The metallic phase, which is driven by correlations from the BI,
displays a peculiar pseudo-gap-like structure in the spectral function
near the Fermi level, as demonstrated in Fig.~\ref{fig:MEM}.
Such features raise the question whether the phase exhibits Fermi-liquid
behavior.

According to the Fermi-liquid theory, the total energy density $\varepsilon$ is
proportional to $T^2$ at low temperatures.
As a relevant check, we calculate the total energy density at various
temperatures and show the results in Fig.~\ref{fig:FL} for various values of $\Delta$.
Indeed $\varepsilon$ appears to be proportional to $T^2$ within
statistical errors for all values of $\Delta$ examined
and we presume that the metallic phase appearing in the IHM is a Fermi liquid.
In addition, we have also computed the imaginary part of the self-energy,
to find that the quasiparticle has an infinite lifetime at the Fermi level;
this is also consistent with the Fermi-liquid picture.

\subsection{Critical point of the Mott transition}

\begin{figure}
	\includegraphics[width=0.4\textwidth]{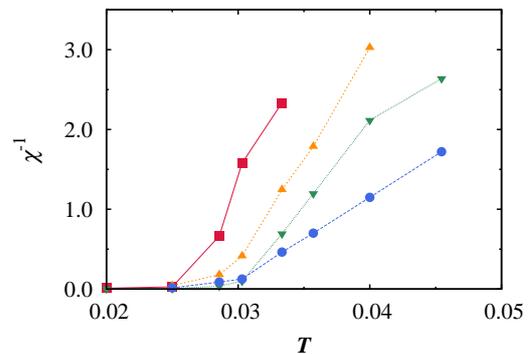}
	\caption{(Color online) Inverse susceptibility $\chi^{-1}$ versus
		temperature $T$ for the staggered lattice potential
		$\Delta = 0.5$ [(red) squares], $1.0$ [(orange) triangles],
	$3.0$ [(green) inverted triangles], and $5.0$ [(blue) circles].
	}
	\label{fig:Tc}
\end{figure}

\begin{figure}
	\includegraphics[width=0.4\textwidth]{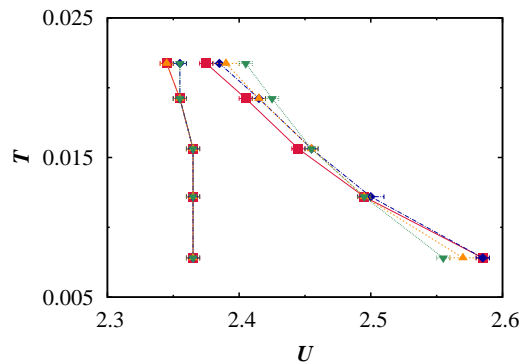}
	\caption{(Color online)
		Coexistence region for $\Delta= 0.5$ (diamonds), $1.0$ (triangles),
        	and $3.0$ (inverted triangles).
		For clear comparison with the Hubbard model [represented by (red) squares],
		data for $\Delta=0.5, 1.0$, and $3.0$ are shifted to the left
        	by the amount $\delta U=0.3$, $1.0$, and $4.69$, respectively.
	}
	\label{fig:coexist}
\end{figure}

In this subsection, we consider how the phase diagram depends on the staggered lattice
potential $\Delta$.
Specifically, we compute the critical temperature $T_c$ of the Mott-Hubbard transition for various
values of $\Delta$.
One way of obtaining $T_c$ is to utilize the divergence of the susceptibility at the critical point.
By analogy with a fluid system,~\cite{Rozenberg1999} we define the susceptibility as
\begin{equation}
\chi \equiv \mathop{\rm Max}_U \left| \frac{\partial d_O}{\partial U} \right|
\end{equation}
at given temperature $T$.
In view of the divergence at the critical point, one can identify the
critical temperature as the temperature where the inverse susceptibility vanishes.
In Fig.~\ref{fig:Tc} we plot the inverse susceptibility $\chi^{-1}$ versus temperature $T$
for several values of $\Delta$.
For given $\Delta$, as the temperature is lowered, the inverse susceptibility decreases and
eventually vanishes, from which the critical temperature can be estimated.
Figure~\ref{fig:Tc} illustrates that the critical temperature generally
increases with the strength of the staggered lattice potential.

We can reach a similar conclusion when we consider the critical interaction
strengths $U_{c1}$ and $U_{c2}$ directly.
As demonstrated in Fig.~\ref{fig:coexist},
variations in $U_{c1}$ with temperature $T$ are rather insensitive to the
value of $\Delta$, while the increase in $\Delta$ suppresses the change in $U_{c2}$
with the temperature.
This implies that the critical point is located at higher temperatures for larger values of $\Delta$.

\subsection{Phase diagram at low temperatures}
\begin{figure}
	\includegraphics[width=0.4\textwidth]{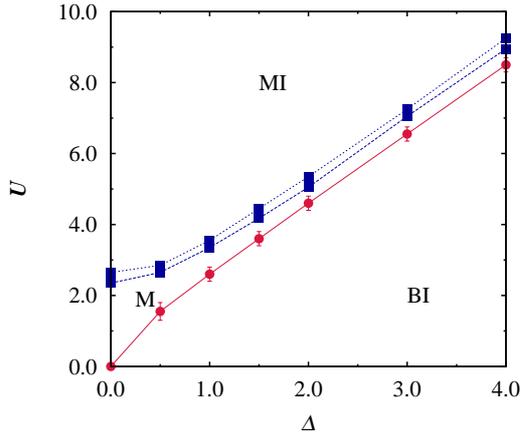}
	\caption{(Color online)
		  Phase diagram on the plane of $\Delta$ and $U$ at temperature
		  $T=1/128$. band insulator (BI), metal (M), and Mott insulator (MI) phases.
		  The (blue) squares represent
		  $U_{c1}$ and $U_{c2}$ of the Mott transition. The (red)
		  circles and vertical bars describe $U_{co}$ and the crossover region between the BI and the metal.
	}
	\label{fig:DUphase}
\end{figure}

Figure~\ref{fig:DUphase} depicts three regions, corresponding to the BI, metal, and MI phases
on the plane of $\Delta$ and $U$ at temperature $T{=}1/128$, which is the lowest temperature examined.
We can observe two prominent differences between the resulting phase diagram and the two zero-temperature phase diagrams obtained in IPT-DMFT studies~\cite{Garg2006,Craco2008}.

First, in our phase diagram the crossover interaction strength $U_{co}$ increases gradually from $0$ as $\Delta$ is turned on.
This is quite in contrast with the rather drastic increase for small
$\Delta$ in Ref.~\onlinecite{Garg2006}. 
Further, here the width of the metallic region apparently remains constant above $\Delta=2$,
which suggests that the metallic phase should extend to high values of $\Delta$.
We have indeed confirmed its existence even for $\Delta=8$.
This qualitatively contradicts the existing prediction that the metallic phase would
cease to exist around $\Delta=1.5$, beyond which a coexistence region
between the BI and the MI develops.~\cite{Craco2008}
At this stage the origin of the discrepancy is not clear and its resolution may  require further study.

\section{summary}
\label{sec:summary}

We have studied the IHM in infinite dimensions by means of the
DMFT combined with the CTQMC method.
The dependence of the double-occupancy and the staggered charge density on the
interaction strength as well as the Fermi-level spectral weight
exhibits crossover behavior from a BI to a metal and, subsequently,
a transition to an MI.
The transition to an MI is of the first order,
and the critical temperature has been found to be higher for stronger
staggered lattice potentials.
Analyzing the temperature dependence of the energy density, we have shown that
the intermediate metallic phase is a Fermi liquid.
Finally, when the staggered lattice potential is strong, this metallic phase has
been found to persist even at very low temperatures.

\begin{acknowledgments}
We thank Prof. P. Werner and Dr. H. Lee for helpful discussions on the CTQMC method and MEM.
This work was supported by the National Research Foundation of Korea
through Grants 
Nos. 2008-0061893 and 2013R1A1A2007959 (A.J.K and G.S.J.), and Grant No. 
2012R1A2A4A01004419 (M.Y.C.).
\end{acknowledgments}

\bibliography{myref}
\end{document}